\documentclass[10pt,onecolumn,a4paper,conference]{IEEEtran}

\usepackage{cite}      

\usepackage{amsmath}   
\usepackage{array}

\usepackage{amsfonts} \usepackage{amssymb}
\usepackage{epsfig}
\usepackage{theorem}
\usepackage[latin1] {inputenc}

\newtheorem{thm}{Theorem}
\newtheorem{defi}{Definition}

\newcommand{\tr}[2][]{\ensuremath{\text{tr}_{#1} \left[#2 \right] }}
\newcommand{\mce}{\ensuremath{{\cal{E}}}}
\newcommand{\mch}{\ensuremath{{\cal{H}}}}
\newcommand{\mcc}{\ensuremath{{\cal{C}}}}

\newcommand{\rl}{\ensuremath{\overline{\rho}}}
\newcommand{\sil}{\ensuremath{\overline{\sigma}}}
\newcommand{\mcs}{\ensuremath{{\cal{S}}}}
\newcommand{\mcx}{\ensuremath{{\cal{X}}}}
\newcommand{\mcp}{\ensuremath{{\cal{P}}}}

\begin{document}

\title{{\huge Zero-error capacity of a quantum channel} \vspace{-0.4cm}}

\author{\authorblockN{Rex A. C. Medeiros and Francisco M. de Assis}
\authorblockA{Federal University of Campina Grande\\
Department of Electrical Engineering\\
Av. Apr\'{i}gio Veloso, 882, Bodocong\'{o}, 58109-970\\
Campina Grande, Para\'{i}ba, Brazil\\
e-mail: [rex,fmarcos]@dee.ufcg.edu.br
}
}



\maketitle


%
\IEEEpeerreviewmaketitle

\vspace{-0.2cm}

The zero-error capacity of a discrete classical channel was first defined by Shannon as the least upper bound of rates for which one transmits information with zero probability of error~[C. Shannon, {\em IRE Trans. Inform. Theory}, IT-2(3):8--19, 1956]. Here, we define the quantum zero-error capacity, a new kind of classical capacity of a noisy quantum channel $\mcc$ represented by a trace-preserving map (TPM) $\mce(\cdot)$. Moreover, the necessary requirement for which a quantum channel has zero-error capacity greater than zero is also given. All proofs will appear in the full paper. Indeed, we will show the connection between quantum zero-error capacity and capacity of a graph.

Let $\mcx$ be the set of possible input states to the quantum channel $\mce(\cdot)$, belonging to a $d$-dimensional Hilbert space $\mch$, and let $\rl \in \mcx$. We denote $\sil_k = \mce(\rl)$ the received quantum state when $\rl$ is transmitted through the quantum channel. Because knowledge of post-measurement states is not important, measurements are performed by means of a POVM (Positive Operator-Valued Measurements) $\{E_j\}$, where $\sum_j E_j =I$. Assume that $p(j|i)$ denotes the probability of Bob measures $j$ given that Alice sent the state $\rl_i$. Then, $ p(j|i) = \tr{\sil_i E_j}$.

We define the zero-error capacity of a quantum channel for product states. A product of any $n$ input states will be called an input quantum codeword, $\rho_i = \rl_{i_1} \otimes \dots \otimes \rl_{i_n}$, belonging to a $d^n$-dimensional Hilbert space $\mch^n$. A mapping of $K$ classical messages (which we may take to be the integers $1,\dots,K$) into a subset of input quantum codewords will be called a quantum block code of length $n$. Thus, $\frac{1}{n}\log K$ will be the rate for this code. A piece of $n$ output indices obtained from measurements performed by means of a POVM $\{E_j\}$ will be called an output word, $w = \{1,\dots,N\}^n$. A decoding scheme for a quantum block code of length $n$ is a function that univocally associates each output word with integers 1 to $K$ representing classical messages. The probability of error for this code is greater than zero if the system identifies a different message from the message sent.

\begin{defi}[Zero-error capacity of a quantum channel]
\label{def:qzec}
Let $\mce$ be a trace-preserving quantum map representing a noisy quantum channel. The zero-error capacity for this channel, denoted by $C^{0}(\mce)$, is the least upper bound of achievable rates with probability of error equal to zero. That is,
\begin{equation}
\label{eq:qzec}
  C^{0}({\mce}) = \sup_{n} \frac{1}{n}\log K(n),
\end{equation}
where $K(n)$ stands for the number of classical messages that the system can transmit without error, when a quantum block code of length $n$ is used.
\end{defi}

According to the classical definition of zero-error capacity, it is possible to establish a condition for which a quantum has quantum zero-error capacity greater than zero.

\begin{thm}
Let $\mcs=\{\rl_i\}$, $\mcs \subset \mcx$ be a set of $M\leq d$ quantum states, and let $\mcp = \{E_j\}$ be a POVM having $N\ge M$ elements such that $\sum_j E_j =I$. Consider the subsets
\begin{equation}
A_k = \{ j \in \{1,\dots,N\}; \; p(j|k) > 0\}; \; k \in \{1,\dots,M\}.
\end{equation}
Then, the quantum channel $\mce$ has zero-error capacity greater than zero iff there exists a set $\mcs$  and a POVM $\mcp$ for which at least one pair  $(a,b) \in \{1, \dots, M\}^2$, $a \neq b$, the subsets $A_a$ and $A_b$ are disjoints, i.e., $A_a \cap A_b = \oslash$.
\end{thm}
The respective quantum states $\rl_a$ and $\rl_b$ are said to be non-adjacent in $\mce(\cdot)$ for the POVM  $\mcp$.

\begin{defi}[Optimum $(\mcs,\mcp)$ for $\mce(\cdot)$]. Consider the set of pairs $A = \{(a_i,b_i) \in \{1, \dots, M\}^2\}$ for which $A_{a_i} \cap A_{b_i} = \oslash$. The optimum $(\mcs,\mcp)$ for $\mce(\cdot)$ is composed of a set $\mcs = \{\rl_i\}$ and a POVM $\mcp = \{E_j\}$ maximizing the cardinality $\parallel A\parallel$.
\end{defi}

\begin{thm}
\label{th:om}
The zero-error capacity of a noisy quantum channel $\mce(\cdot)$ is achieved iff the optimum $(\mcs,\mcp)$ is used. 
\end{thm}

At first glance, one may think the Definition~\ref{def:qzec} is identical to the classical case. However, according with Theorem~\ref{th:om}, it turns out that the supremum in Eq.~(\ref{eq:qzec}) is taken over all sets $\mcs$ of possible input states $\rl$ for the quantum channel $\mcc$ and all POVM $\{E_j\}$. The restriction on the creation of quantum states means that each state $\rl$ can not be entangled with other states across two or more uses of the channel. However, we point that, likewise Holevo-Schumacher-Westmoreland's capacity, the decoding scheme allows measurements entangled across $n$ uses of the channel.

\end{document}